\documentclass[
    ,final            
  ]
  {aipproc}

\layoutstyle{6x9}

\usepackage{xspace}
\def\bbbar {\ensuremath{b\overline b}\xspace}
\def\B       {\ensuremath{B}\xspace}
\def\Bs      {\ensuremath{B_s^0}\xspace}
\def\Bd      {\ensuremath{B^0}\xspace}
\def\jpsi     {\ensuremath{{J\mskip -3mu/\mskip -2mu\psi\mskip      2mu}}\xspace}

\begin{document}

\title{Recent LHCb Results}



\author{Giacomo Graziani, on behalf of the LHCb collaboration}{
  address={INFN, Sezione di Firenze (Italy)}
}

\begin{abstract}
 The LHCb experiment started its physics program with
 the  37 pb$^{-1}$ of pp collisions at $\sqrt{s}$=7 TeV delivered  by the
 LHC during 2010.
The performances and capability of the experiment, conceived for precision measurements  in the heavy flavour sector,
are illustrated through the first results from the experimental core
program. A rich set of production studies 
provide precision QCD and EW tests in the unique high rapidity region
covered by LHCb. Notably, results for  W and Z production are very
encouraging for setting constraints on the parton PDFs. 
\end{abstract}

\maketitle


\section{The Concept of LHCb}

\begin{figure}[b]
  \includegraphics[width=.9\textwidth]{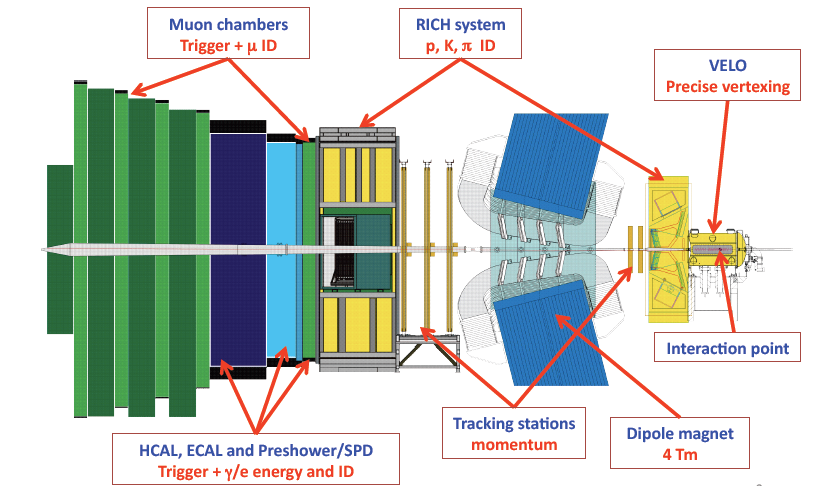}
  \caption{The LHCb detector}
  \label{fig:lhcbDetector}
\end{figure}
The LHCb experiment\cite{bib:lhcbDetector} is searching for new physics through non standard
virtual contributions to CP violating and rare decays of heavy
hadrons. It is implemented as a single arm spectrometer covering the
pseudorapidity region $1.9 < \eta < 4.9$, where the largest production rate of \bbbar 
pairs per solid angle is expected at LHC energies. This coverage is unique at the LHC
and complementary to the general purpose detectors ATLAS and CMS. The
detector design, sketched in fig.~\ref{fig:lhcbDetector}, emphasizes proper time resolution, to identify the short
lived $b$ hadrons and resolve the fast \Bs oscillations, while
excellent invariant mass resolution and particle identification (PID) capabilities are
needed to disentangle decays of heavy flavours from the dominant 
hadronic background. 
These detector features also make LHCb a powerful tool for heavy flavour spectroscopy
and production studies in its unique rapidity
range, providing precision QCD tests at the unprecedented energies
reached by the LHC.

The geometry of the Vertex Locator (VELO) is
optimized for detecting the decays of $b$ hadrons in the forward direction.
Its Si $\mu$-strips sensors are perpendicular to the beam  and are located on two
retractable half stations. When stable beam is declared, the two
halves are closed and the inner border of the active area is only 8 mm
away from the beam axis. Tracking is completed
by a set of stations equipped with Si $\mu$-strips (for the inner part) and straw tubes (for the
outer part), located upstream and downstream a warm dipole magnet
providing an integrated field of 4 Tm. The combined performances of two RICH detectors result in
excellent $\pi$/K/p separation in the 1--100 GeV/c momentum range. The
calorimetric system provides e/$\gamma$/hadrons separation through a scintillating pad detector, a
preshower, electromagnetic (shashlik) and hadronic (Fe/scint. tiles)
calorimeters. Muons are identified by five stations equipped with
multi--wire proportional chambers (or GEM chambers for the most inner part),
interspaced by iron absobsers.
A fast, flexible and efficient trigger is
essential to extract the interesting events from the nominal 40 MHz collision
rate. It is implemented in two levels: the initial decision comes from
an hardware level (L0),
selecting particles of high transverse momentum using the informations
from the calorimetric and muon systems, which can operate at 40 MHz with a
relatively low $p_T$ threshold ($\sim$ 1 
GeV/c). A software level (HLT), running on a massive computer farm,
performs an online event reconstruction, reducing the event rate from
1 MHz to 2--3 kHz through inclusive and exclusive selections.

The experiment was designed to work at a luminosity of $2\times
10^{32}$ cm$^{-2}$s$^{-1}$, 50 times lower than the LHC design luminosity,
in order to minimize the pile--up.

\section{The 2010 run}
During its spectacular startup in 2010, the LHC machine gradually increased
luminosity over 5 orders of magnitude, reaching the
nominal LHCb value already by the end of the run. Due to the
limited number of colliding bunches in this phase, LHCb acquired
events with up to 2.6 visible collisions 
per crossing, six times the design value. The flexibility of trigger
and DAQ systems allowed the experiment to cope with this high pile--up while keeping
high efficiency for key channels, thereby maintaining
the statistics for physics studies. The detector was fully
working, with a negligible amount of dead channels in all subsystems,
and the overall DAQ efficiency was 90\%.  
Eventually, LHCb recorded an integrated luminosity of 37 pb$^{-1}$,
similar to ATLAS and CMS. 
Though being a small sample compared to the
nominal 2 fb$^{-1}$ per year expected in the longer term, this dataset
corresponds already to about 5 billion $b$ events in the detector
acceptance, sufficient statistic for the first competitive physics results, as
well as the validation of detector performances. We will present in this
contribution a selection of results from $b$ decays,
illustrating the potentiality of the experiment, followed by an
overview of QCD and EW studies at LHCb.

\section{Proper Time Reconstruction and the \Bs}

The VELO vertex resolution was measured on real data by 
comparing the reconstructed vertexes from randomly chosen subgroups of
tracks in the same primary vertex. For a typical vertex of 25
tracks, a resolution of 16 (76) $\mu$m was obtained for the transverse
(longitudinal) direction. The impact parameter resolution was measured
to be 13 + (26 GeV/c)/$p_T$ $\mu$m, corresponding to a typical
proper time resolution for \B decays of 50 fs. 
Preliminary results(\cite{bib:confnotes}, LHCb-CONF-2011-001) for the lifetime measurements of $B^0, B^+,  B^0_s$  
and $\Lambda_b$, using final states containing a \jpsi, are shown on
table \ref{tab:lifetimes}. 
The excellent agreement with PDG averages
demonstrates the level of control on the lifetime scale.

\begin{table}[hbt]
\begin{tabular}{lrll}
\hline
Channel                 & event yield    & lifetime (ps), with stat. and syst. error  & PDG2010 (ps) \\ 
\hline
$B^+ \to \jpsi K^+$    & 6741 $\pm$ 85  & 1.689 $\pm$ 0.022 $\pm$ 0.047 & 1.638 $\pm$ 0.011 \\
$B^0 \to \jpsi K^{*0}$ & 2668 $\pm$ 58  & 1.512 $\pm$ 0.032 $\pm$ 0.042 & 1.525 $\pm$ 0.009 \\
$B^0 \to \jpsi K_S$    &  838 $\pm$ 31  & 1.558 $\pm$ 0.056 $\pm$ 0.022 & 1.525 $\pm$ 0.009 \\
$\Bs \to \jpsi \phi$   &  570 $\pm$ 24  & 1.447 $\pm$ 0.064 $\pm$ 0.056 & 1.477 $\pm$ 0.046 \\
$\Lambda_b \to \jpsi \Lambda$    
                        &  187 $\pm$ 16  & 1.353 $\pm$ 0.108 $\pm$ 0.035 & 1.391$^{+0.038}_{-0.037}$ \\
\hline
\end{tabular}
\caption{Event yields and lifetimes obtained from several $b$ hadron decays
  to exclusive states containing a \jpsi  in the range 0.3 $<$ t $<$ 14
  ps. Events are triggered using dimuon candidates with invariant mass
  compatible with the \jpsi and minimum $p_T$ of 500 MeV/c, without any bias on lifetime.
  The current PDG values for the lifetimes are shown for comparison.}
\label{tab:lifetimes}
\end{table}

\begin{figure}[b]
  \includegraphics[width=\textwidth]{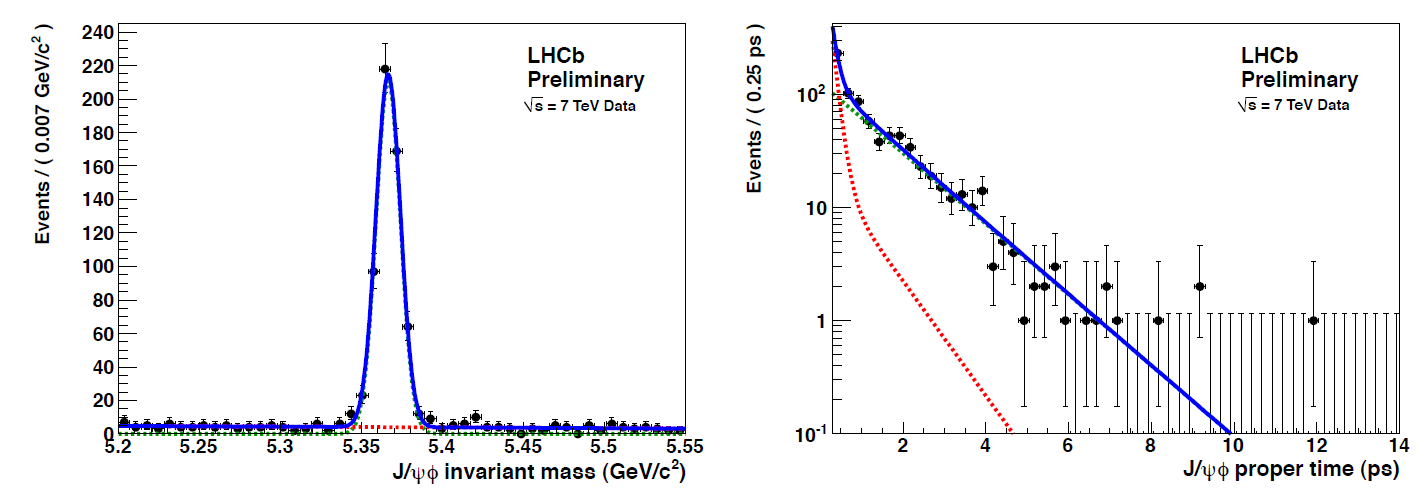}
  \caption{Invariant mass (left) and proper time t (right) distributions
    for the $\Bs \to \jpsi \phi$ candidates with t>0.3 ps. The
    signal (green dashed line), background (red dashed line) and total
    (blue solid line) contributions obtained from a two-dimensional fit are shown on
    the plots. } 
  \label{fig:jpsiPhi}
\end{figure}

\Bs mesons, largely unexplored at the B factories, provide some of
the most promising channels for the emergence of new physics. From
the ``golden mode'' $\Bs \to \jpsi \phi$
the CP--violating oscillation phase $\phi_s$, for which an accurate
(error $\sim$ 2 mrad) SM prediction exists, can be
measured. Figure~\ref{fig:jpsiPhi} shows the invariant 
mass and lifetime distributions for the candidate events from lifetime
unbiased triggers. The combination of impact parameter, invariant
mass resolution (7 MeV/c$^{2}$, with the \jpsi mass constrained to
the PDG value) and PID performances results in a very clean
event sample. As can be seen from table~\ref{tab:lifetimes}, despite
the limited statistics, the accuracy of the lifetime
measurement in this mode is already comparable with the Tevatron
results, obtained from $\sim$ 100 times more integrated luminosity.

\begin{figure}[b]
  \includegraphics[width=\textwidth]{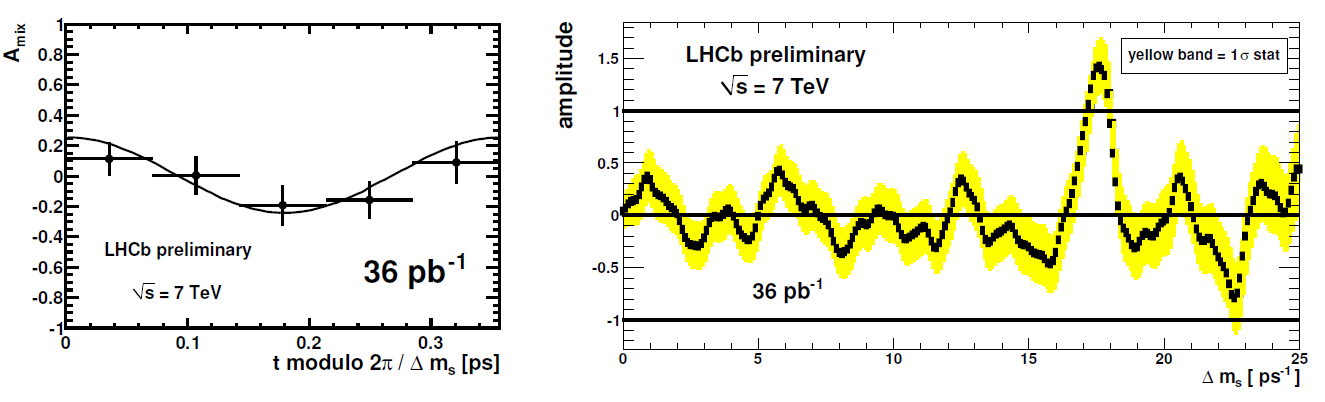}
  \caption{ Preliminary results of the \Bs oscillations analysis. On the left: the flavour asymmetry for \Bs
candidates as a function of proper time modulo 2$\pi/\Delta m_s$.
The fitted asymmetry is superimposed. On the right: fitted oscillation
amplitude as a function of $\Delta m_s$.} 
  \label{fig:b0soscilla}
\end{figure}

Flavour oscillations of \Bs mesons were searched for in the more abundant $D_s \pi$ and
$D_s 3\pi$ modes.
Neural network based tagging algorithm were developed and calibrated
on real data using self--tagging modes. The limited statistics of
the control modes is presently the main source of uncertainty and prevents
the use of the powerful same side tagger for \Bs with this dataset. However, a tagging
power of $3.8 \pm 2.1 \%$ was obtained, allowing  \Bs
oscillations to be observed, with a frequency measured to be(\cite{bib:confnotes},
LHCb-CONF-2011-005):
$$ \Delta m_s = 17.63 \pm 0.11 (stat) \pm 0.04 (syst) ~ \textrm{ps}^{-1}$$
as illustrated in figure~\ref{fig:b0soscilla}. The measurement
agrees well with the world's best published result 
to date from CDF
($17.77 \pm 0.10 \pm 0.07 ~ \textrm{ps}^{-1}$).

A time--dependent tagged analysis was performed for $\Bs \to \jpsi
\phi$, leading to a first loose bound on
$\phi_s$~(\cite{bib:confnotes}, LHCb-CONF-2011-006). We expect to reach an accuracy of 35
mrad, comparable to the SM predicted value, with  the first inverse fb
of data expected in 2011.

Another key \Bs channel is the ultra-rare $\mu^+\mu^-$ decay, with a
predicted BR of $(3.2 \pm 0.2)\cdot 10^{-9}$ in the SM, that could receive
significant contributions from new pseudo--scalar intermediate states. 
The channel is also very clean experimentally, with no significant
peaking background. The analysis of the full 37 pb$^{-1}$ data sample showed no
candidate events, leading to a 95\% CL limit of~\cite{bib:Aaij:2011rj}
$$ BR(\Bs \to \mu^+\mu^-) < 5.6 \cdot 10^{-8}$$
which is very close to the best limit from CDF ($4.3 \cdot 10^{-8}$),
obtained from 3.7 fb$^{-1}$. The sensitivity is expected to attain the SM level
in LHCb with about 2 fb$^{-1}$.

The competitiveness of the experiment for \Bs physics is also
demonstrated by the first observation of several decay channels:
$\Bs \to D_{s2}X\mu\nu$, $\Bs \to K^{*0}\overline{K^{*0}}$, $\Bs \to D^0K^{*}$,
$\Bs \to \jpsi f_0$~\cite{bib:Aaij:2011ju, bib:Aaij:2011fx, bib:confnotes}. The latter, a CP eigenstate, can provide an
interesting contribution to the measurement of $\phi_s$.

\section{Particle ID at work}
The most spectacular demonstration of the RICH's performances is
provided by the analysis of charmless $\B \to hh$
decays, where $\pi/K/p$ discrimination
is essential to disentangle the $B^0\to\pi^+\pi^-, B^0\to K^+\pi^-,
\Bs\to K^+K^-, \Bs\to  \pi^+K^-, \Lambda_b\to p K^-$ and $\Lambda_b \to
p \pi^-$ modes. PID performances are calibrated on real data
using the abundant $D^*$ and $\Lambda$ decays. The $\B \to hh$ yields are then
fitted simultaneously taking into 
account the expected cross feeds. The $\Bd\to\pi^+\pi^-$ and $\Bs\to
K^+K^-$ modes will allow for a measurement of the CKM angle $\gamma$
from loop diagrams. From the 2010 data sample we extract promising
yields of 275 $\pm$ 24 $\Bd\to\pi^+\pi^-$ and 333 $\pm$ 21 $\Bs\to
K^+K^-$ candidates, corresponding to about one fourth of 
what obtained by CDF from 1 fb$^{-1}$.

\begin{figure}[b]
  \includegraphics[width=\textwidth]{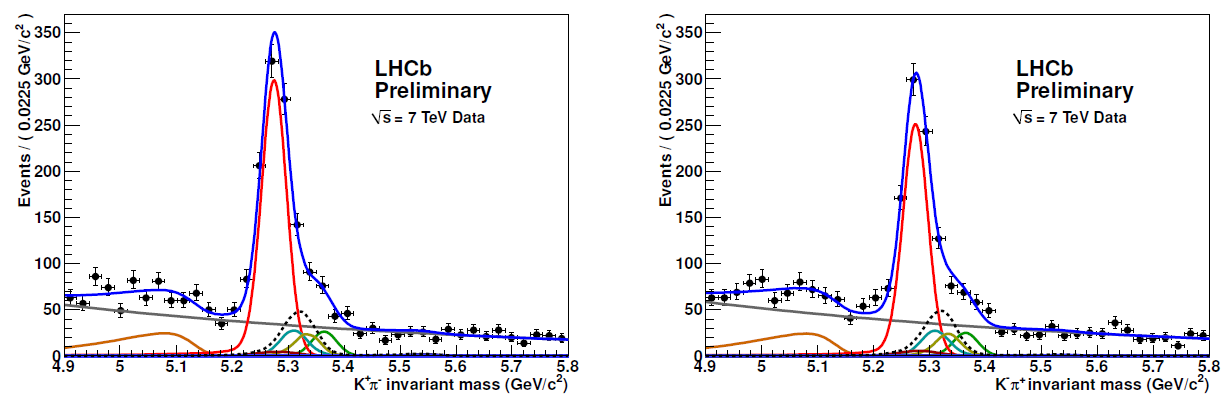}
  \caption{Invariant mass distributions for (left) $B^0 \to K^+\pi^-$
    and (right)  $\overline{\Bd} \to K^-\pi^+$ candidates. The dots
    represent the data, while the curves show the result
    of the unbinned ML analysis fit: total (blue), signal (red),
    combinatorial background (gray) and cross--feed components
  from the other $B\to hh$ modes.} 
  \label{fig:Kpi}
\end{figure}

Although more statistics is needed for a competitive $\gamma$ measurement,
with 2010 data we already reached the sensitivity to confirm the direct CP
violation in the $\Bd \to K^+\pi^-$ mode from the time integrated CP
asymmetry, as shown in figure~\ref{fig:Kpi}. The possible biases on
the measured raw asymmetry were constrained from real data: the
detector asymmetry was found to be $A_D = -0.004 \pm 0.004$ from 
$D, D^{*} \to K\pi$ decays,
while the production asymmetry was estimated to be  $A_p = 0.009 \pm
0.008$ from $B^{\pm}\to \jpsi K^{\pm}$. The results, compared with
world averages from HFAG, are~(\cite{bib:confnotes}, LHCb-CONF-2011-023):
\medskip

\begin{tabular}{lcc}
mode    & LHCb (preliminary) & HFAG average \\ \hline
$A_{CP}(\Bd \to K^+\pi^-)$  &  ~~~~~$-0.077 \pm 0.033 \pm 0.007$~~~~~ & $-0.098^{+0.012}_{-0.011}$ \\   
$A_{CP}(\Bs \to \pi^+K^-)$  &  $0.15 \pm 0.19 \pm 0.02$ & $0.39 \pm 0.17$
\end{tabular}
\medskip

The 2.3$\sigma$ effect for \Bd is the first hint for CP violation
measured at LHC. Also in this case, the measurement for \Bs is
already competitive with the world average.

\medskip
Performances of the calorimeters were also excellent, with a
resolution of 7.2 MeV/c$^2$ for the $\pi^0$ mass from unconverted
photons and a response uniformity within 2\%. Despite the harsh
hadronic environment and 
the high pile--up experienced in 2010, clean samples of exclusive
radiative decays as $B^0\to K^*\gamma$ and $\chi_c\to \jpsi\gamma$
could be reconstructed, implying nice prospects for the physics
with the rare $b\to s\gamma$ modes.

\section{LHCb as a general purpose forward detector}
Particle production rates can be studied by LHCb in a
rapidity range complementary to the general purpose detectors. The
excellent PID and vertexing capabilities, combined with the low $p_T$
thresholds of the trigger,  allow to select a
wide range of exclusive  states with good efficiency. LHCb can
contribute in particular to heavy flavour 
spectroscopy, studying of X(3872) and the other unexpected
states recently observed, the $B_c$ and the other double heavy flavour
states.
We summarize here some of these studies, that are discussed in more details
in the other LHCb contributions to these Proceedings.

\begin{figure}[b]
\begin{minipage}{.55\textwidth}
  \includegraphics[width=\textwidth]{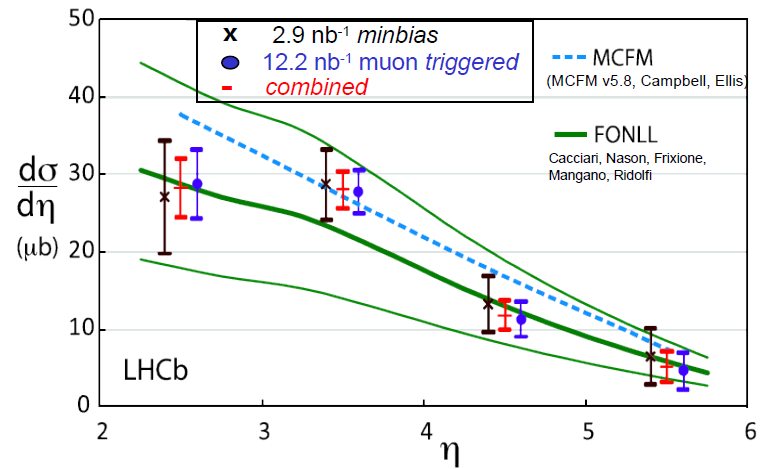}
\end{minipage}
\begin{minipage}{.45\textwidth}
  \includegraphics[width=.97\textwidth]{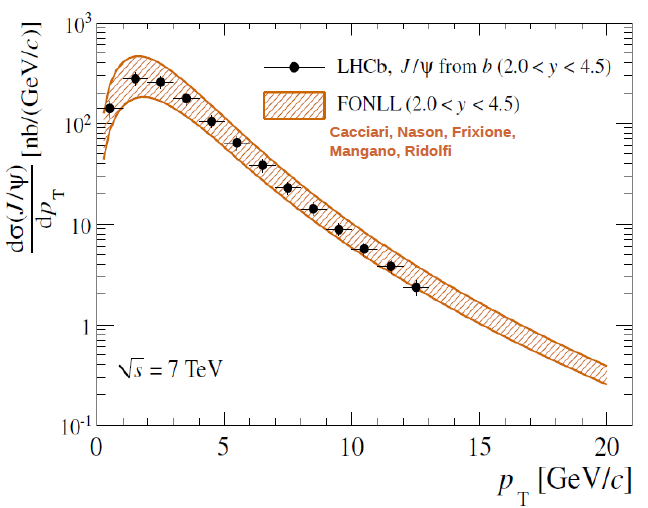}
\end{minipage}
  \caption{Results of \bbbar cross section measurements from
charm events (left, as a function of pseudorapidity) and from delayed
\jpsi (right, as a function of the  \jpsi transverse momentum),
compared to theoretical predictions.} 
\label{fig:bbprod}
\end{figure}

The production cross section of $b$ hadrons is of obvious importance
for LHCb, but also as a QCD testbench and as a crucial input for
background determination in many key channels at LHC, notably the
search for the Higgs boson. It was measured  using the inclusive $b\to
D^0X\mu^-\nu$ mode from just the first 15 nb$^{-1}$ of
data~\cite{bib:Aaij:2010gn}. A measurement of similar accuracy was possible with 5
pb$^{-1}$ using events with a delayed \jpsi decaying to $\mu^+\mu^-$~\cite{bib:Aaij:2011jh}. 
The results, compared to some theoretical predictions, are showed in fig.~\ref{fig:bbprod}. 
Extrapolating to the full rapidity range, we get 
\begin{eqnarray*}
 \sigma(p\overline{p}\to \bbbar X)|_{\textrm{s=7 TeV}}  & =  (284 \pm 20 \pm 49) ~ \mu  b ~~~~~~  & (b\to D^0X\mu^-\nu)\\ 
                                                              & =  (288 \pm ~~4 \pm 48) ~ \mu  b ~~~~~~  & (b\to \jpsi X),
\end{eqnarray*}
in good agreement with the production models tuned on the Tevatron results.

The first low--luminosity data allowed LHCb to allocate a relevant fraction of
the trigger bandwidth to charm physics. Production of all open charm states
was measured using minimum--bias trigger, resulting in valuable QCD tests
(notably from the $D^+/D^+_s$ production ratio) and reducing
uncertainties for backgrounds in many CPV measurements. 

\begin{figure}[t]
\begin{minipage}{.43\textwidth}
\includegraphics[width=.9\textwidth]{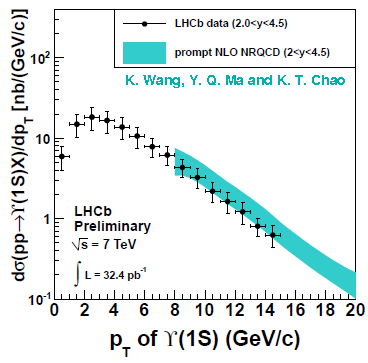}
\end{minipage}
\begin{minipage}{.57\textwidth}
\includegraphics[width=.95\textwidth]{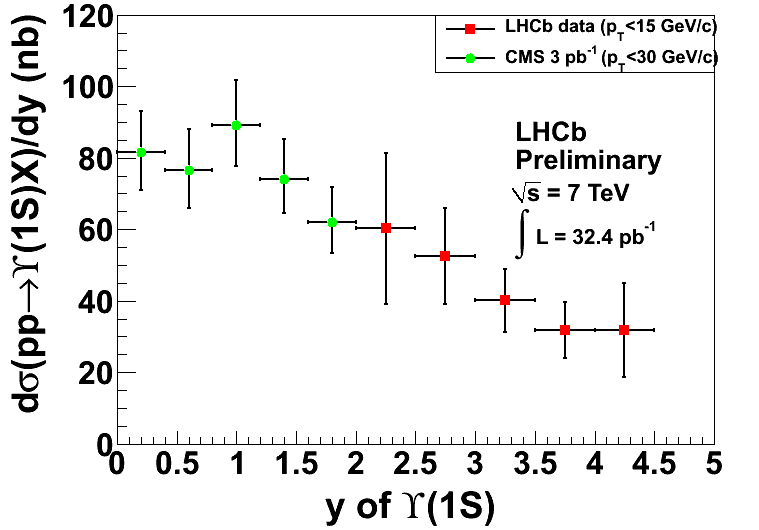}
\end{minipage}
  \caption{The cross section for $\Upsilon$(1S) production
    is shown as a function of transverse momentum
    (left, compared with theory) and pseudorapidity (right, compared
    to CMS in the central region).}
\label{fig:upsilon}
\end{figure}

A rich quarkonium physics program has begun by measuring the
production rates of prompt \jpsi, $\psi(2S)$, $\chi_c$,
$\Upsilon$(1S). 
Figure~\ref{fig:upsilon} shows the results for bottomonium,
compared with the CMS values at lower $\eta$, nicely illustrating the
complementarity  of LHC experiments. Data also cover production at very low transverse
momentum, down to less than 1 GeV/c, well below  the typical threshold
of theoretical predictions! 

Double \jpsi production was also
observed, as well as $B_c$ production in the $\jpsi \pi^+$
channel~(\cite{bib:confnotes}, LHCb-CONF-2011-009 and LHCb-CONF-2011-017), paving the way for the exploration of states from double
heavy flavour production. 

\section{Scrutinizing parton PDFs}
The LHCb rapidity region allows the study of deep inelastic scattering 
in the unique unexplored high--$Q^2$ regions at very low $x$.
Drell--Yan production of muon pairs can be observed for $Q^2 >5$ 
GeV$^2$, probing parton PDFs down to $x \sim 10^{-6}$.  Production of
$W$ and $Z$ bosons, corresponding to $x\sim 10^{-4}, Q^2=M_{W,Z}^2$, 
has been observed already with the first 17 pb$^{-1}$ of 
data~(\cite{bib:confnotes}, LHCb-CONF-2011-012).

The cross sections are predicted with accuracy varying
with $\eta$ from 3 to 10\%, the uncertainty on parton PDFs
contributing to a large extent.
Constraints to the PDFs can thus be obtained, particularly from
the charge asymmetry of $W$ production, since many systematics cancel
and its $\eta$ dependence is very sensitive on PDFs.

\begin{figure}[t]
\begin{minipage}{.5\textwidth}
\includegraphics[width=\textwidth]{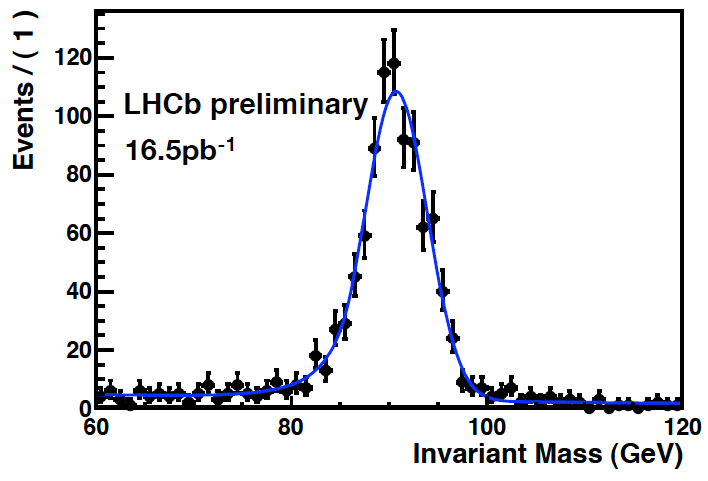}
\end{minipage}
\begin{minipage}{.5\textwidth}
\includegraphics[width=\textwidth]{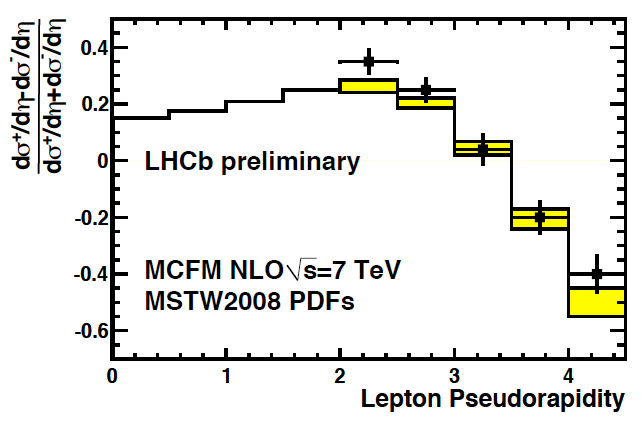}
\end{minipage}
  \caption{On the left, invariant mass distribution for the
    $Z\to\mu^+\mu^-$ candidates. On the right, charge asymmetry of the
    $W$ production as a function of $\eta_{\mu}$, compared
    with the prediction of the NLO MCFM generator. The shaded area shows the
    uncertainty due to the MSTW08 PDF set used in the model.}
\label{fig:zw}
\end{figure}

A clean sample of $Z\to\mu^+\mu^-$ decays can be easily selected using
the invariant mass constraint on muons having  $p_T>$ 20 GeV/c,
with almost no background expected. With an estimated 69\% efficiency, 833 candidate
events survive the selection (see fig.~\ref{fig:zw}).

The signature for $W$ decays is momentum imbalance in the
transverse plane, with an isolated high--$p_T$ ($>$ 20 GeV/c) muon and little other
activity in the event to suppress background from QCD jets. The muon
is also required to be compatible with the primary vertex to suppress backgrounds
from $b, c$ decays. For the resulting 7624 $W^+$ and
5732 $W^-$ candidates we estimate a selection efficiency of 30\% and a
signal to background ratio of about 1.6. The resulting cross sections are
\begin{eqnarray*}
\sigma_Z ( {\scriptstyle 81<m_Z<101 \rm{~GeV/c}^2}) \times BR(Z\to\mu^+\mu^-,
{\scriptstyle 2<\eta_{\mu}<4.5, p_{T\mu} > 20 \rm{~GeV/c}}) = & 
73 \pm 4 \pm 7 \textrm{~pb} \\
\sigma_{W^+} \times BR(W\to\mu\nu, {\scriptstyle 2<\eta_{\mu}<4.5, p_{T\mu} > 20 \rm{~GeV/c}}) = & 1007 \pm 48 \pm 100 \textrm{~pb} \\   
\sigma_{W^-} \times BR(W\to\mu\nu, {\scriptstyle 2<\eta_{\mu}<4.5, p_{T\mu} > 20 \rm{~GeV/c}}) = & ~682 \pm 40 \pm ~68 \textrm{~pb} 
\end{eqnarray*}
The systematic errors are dominated by the
uncertainty on luminosity, a component which cancels in
the ratios. As shown in fig.~\ref{fig:zw}, the error on the
$W$  charge asymmetry is already comparable to the
uncertainty due to the PDFs. With the much larger statistics expected for 2011, 
LHCb will start contributing to the determination of the
proton structure. 

\section{Conclusions}
The first data sample acquired in 2010 provided 
a proof of principle of the LHCb concept.
With only 37 pb$^{-1}$, all the
necessary ingredients for the key \B physics measurements, namely
proper time resolution, background suppression and tagging capabilities,
have been demonstrated. The first world class results show that LHCb is
already picking up the baton  from the successful Tevatron \B
physics program, confirming how high--precision measurements in the
heavy flavour sector can be achieved at hadron colliders.

LHCb physics is also expanding well
beyond the core program, with many original results 
on production studies in the unique forward region
covered by the experiment, as documented by the other 8 LHCb contributions
to this conference.

According to the LHC schedule,  we expect to collect our first
fb$^{-1}$ within the 2011 run. A wide range of unexplored flavour
territories is opening out in front of us. 
\bibliographystyle{aipproc}   
\bibliography{graziani}

\begin{thebibliography}{7}
\expandafter\ifx\csname natexlab\endcsname\relax\def\natexlab#1{#1}\fi
\providecommand{\enquote}[1]{``#1''}
\expandafter\ifx\csname url\endcsname\relax
  \def\url#1{\texttt{#1}}\fi
\expandafter\ifx\csname urlprefix\endcsname\relax\def\urlprefix{URL }\fi
\providecommand{\eprint}[2][]{\url{#2}}

\bibitem[Alves et~al.(2008)]{bib:lhcbDetector}
A.~Alves, et~al., \emph{JINST} \textbf{3}, S08005 (2008).

\bibitem[{LHCb preliminary results}(2011)]{bib:confnotes}
{LHCb preliminary results}  (2011), {reports submitted to conferences are
  available on "{\it http://cdsweb.cern.ch/collection/LHCb Conference
  Contributions}"}.

\bibitem[Aaij et~al.(2011{\natexlab{a}})]{bib:Aaij:2011rj}
R.~Aaij, et~al., \emph{Phys.Lett.} \textbf{B699}, 330--340
  (2011{\natexlab{a}}), \eprint{1103.2465}.

\bibitem[Aaij et~al.(2011{\natexlab{b}})]{bib:Aaij:2011ju}
R.~Aaij, et~al., \emph{Phys.Lett.} \textbf{B698}, 14--20 (2011{\natexlab{b}}),
  \eprint{1102.0348}.

\bibitem[Aaij et~al.(2011{\natexlab{c}})]{bib:Aaij:2011fx}
R.~Aaij, et~al., \emph{Phys.Lett.} \textbf{B698}, 115--122
  (2011{\natexlab{c}}), \eprint{1102.0206}.

\bibitem[Aaij et~al.(2010)]{bib:Aaij:2010gn}
R.~Aaij, et~al., \emph{Phys.Lett.} \textbf{B694}, 209--216 (2010),
  \eprint{1009.2731}.

\bibitem[Aaij et~al.(2011{\natexlab{d}})]{bib:Aaij:2011jh}
R.~Aaij, et~al., \emph{Eur.Phys.J.} \textbf{C71}, 1645 (2011{\natexlab{d}}),
  \eprint{1103.0423}.

\end{thebibliography}

\IfFileExists{\jobname.bbl}{}
 {\typeout{}
  \typeout{******************************************}
  \typeout{** Please run "bibtex \jobname" to obtain}
  \typeout{** the bibliography and then re-run LaTeX}
  \typeout{** twice to fix the references!}
  \typeout{******************************************}
  \typeout{}
 }

\end{document}